\documentclass[journal=jacsat,manuscript=article]{achemso}
\usepackage[version=3]{mhchem}

\usepackage{graphicx}  % needed for figures
\usepackage{dcolumn}   % needed for some tables
\usepackage{bm}        % for math
\usepackage{amssymb}   % for mathv
\usepackage{color}
\usepackage{epsfig}
\usepackage{epstopdf}

\author{Anatolii V. Mokshin}
\email{anatolii.mokshin@mail.ru} \affiliation[Kazan Federal
University]{Department of Physics, Kazan Federal University,
Kazan, Russia}

\author{Bulat N. Galimzyanov}
\affiliation[Kazan Federal University]{Department of Physics,
Kazan Federal University, Kazan, Russia}

\title[\texttt{achemso}]
{Steady-State Homogeneous Nucleation and \\ Growth of Water Droplets:\\
Extended Numerical Treatment}

\begin{document}

\begin{abstract}
The steady-state homogeneous vapor-to-liquid nucleation and the
succeeding liquid droplet growth process are studied for water
system by means of the coarse-grained molecular dynamics
simulations with the mW-model suggested originally in [Molinero,
V.; Moore, E. B. \textit{J. Phys. Chem. B} \textbf{2009},
\textit{113}, 4008-4016]. The investigation covers the temperature
range $273 \leq T/K \leq 363$ and the system's pressure $p\simeq
1$~atm. The thermodynamic integration scheme and the extended mean
first passage time method as a tool to find the nucleation and
cluster growth characteristics are applied. The surface tension is
numerically estimated and is compared with the experimental data
for the considered temperature range. We extract the nucleation
characteristics such as the steady-state nucleation rate, the
critical cluster size, the nucleation barrier, the Zeldovich
factor; perform the comparison with the other simulation results
and test the treatment of the simulation results within the
classical nucleation theory. We found that the liquid droplet
growth is unsteady and follows the power law. At that, the growth
laws exhibit the features unified for all the considered
temperatures. The geometry of the nucleated droplets is also
studied.
\end{abstract}

%\\pacs{64.70.Pf, 05.70.Ln, 83.50.-v}

%\\keywords{Nucleation, water, vapor-to-liquid phase transition,
%\droplet growth, mean first passage time method, surface tension,
%\nucleation rate, growth rate}

%\\maketitle
\section{Introduction}

Nucleation is a fundamental process, which characterizes the
mechanisms of the emergence of a new phase, and this is one of the
most widespread ways, by which the phase transitions are
initiated. Although a variety of theoretical descriptions for the
nucleation exists, all of them are based on the same key idea: new
phase starts to evolve within a mother phase from the nuclei, when
they achieve such sizes and shapes, which facilitate the further
stable growth of these nuclei. According to the classical
nucleation theory (CNT), the stability of the nuclei is resulted
by the confrontation of surface and bulk contributions in a free
energy. This is relevant for the homogeneous scenario, which
implies the equal probability for the appearance of a nucleation
event over the sample, as well as for the heterogeneous scenario,
where the some places in a sample are more attractive for the
nucleation events (due to impurities, walls, \emph{etc.}).

Concerning the specific case of the homogeneous droplet nucleation
during the vapor-to-liquid transition in water, there is the
comprehensive experimental material due to series of
investigations (see, for example
\cite{Wolk_JCPB_2001,Luijten_JCP_1997,Mikheev_JCP_2002,Brus_JCP_2008,Brus_JCP_2009,Manka_JCP_2010,Kim_JPCA_2004,Heist_Exp_water,Viisanen_JCP_1993}
and references therein). Here, the direct comparison of the
experimental results with the predictions of the nucleation
theories as well as with the data of the numerical simulations
performed by means of molecular dynamics (MD)
\cite{Matsubara_JCP_2007,Yasuoka/Matsumoto_JCP_1993} and Monte
Carlo \cite{Chen_JCPA_2005,Merikanto_JCP_2004} methods has
revealed the noticeable discrepancies. So, for example, for the
vapor-to-liquid nucleation in water at the identical conditions
(pressure/supersaturation, temperature) the experiments, the
theoretical models (CNT and others) and the numerical simulations
yield the values of the steady-state nucleation rate $J_s$, which
differ by orders of magnitude. Under these circumstances, it could
be quite reasonable to consider the features of the
nucleation-growth kinetics in water at the molecular level,
treating the vapor-to-liquid transition in the context of
molecular interactions and movements.

Recently, Molinero and Moore have suggested a coarse-grained
``monatomic'' model of water (mW), in which the anisotropy in the
molecular interactions is simply realizing by means of an
angular-dependent contribution~\cite{Molinero_JPCB_2009}. The
removal of the atomic interactions from the consideration
accelerates the computations and, thereby, it inspires to probe
the microscopic properties of the system on the extended time
scales. Here, the phase transitions are convenient candidates to
be taken in handling. So, the homogeneous nucleation of ice was
studied within the mW-model of water in
Refs.~\cite{Reinhardt_JCP_2012,Moore_Nature_2011}. Therefore, it
is tempting to extend these studies and to consider the details of
the vapor-to-liquid phase transition on the basis of the mW-model.
An important point is that the mW-model reproduces correctly the
equation of state for the temperature range $250 < T/K < 350$ at
$p\simeq1$\;atm. (see Fig.~4 in Ref.~\cite{Molinero_JPCB_2009}),
which is relevant at the consideration of the droplet nucleation
in water.

From viewpoint of the CNT, three principal parameters are enough
to restore the basic aspects of the steady-state nucleation. These
can be, for example, the steady-state nucleation rate $J_s$, the
nucleation barrier $\Delta G$ and the Zeldovich factor $Z$. Of
course, those three parameters can be taken in another combination
(for example, the ``reduced moment'', the lag-time, and the
steady-state nucleation rate, like it is suggested in
Ref.~\cite{Bartell_JPCB_2004}). Nevertheless, the surface tension
$\sigma$, which characterizes the interphase layer and contributes
to the nucleation barrier through the surface free energy term,
requires the independent treatment~\cite{Mokshin_JPCM_2007}. In
the direct computer simulations, the different adapted convenient
approaches based on the Fowler formula, the Kirkwood-Buff formula
and others are utilizing to define accurately the surface tension
\cite{Horse_PRE_2012,Evans_JPA_1972}. However, there is a
necessity at the study of nucleation to apply such a method (i)
that gives a possibility to estimate the surface tension from the
raw simulation data, (ii) that is applicable to characterize the
surfaces of the microscopical nuclei with a pronounced inherent
curvature, and (iii) that considers the genuine interphase
(vapor-liquid) properties without reference to a vacuum phase.

In the present work, we study the nucleation-growth processes of
water droplets on the basis of MD simulations with the mW-model.
To define the parameters of the nucleation and the droplet growth,
we apply the statistical treatment of the simulation data on the
basis of the thermodynamic integration scheme and the mean first
passage time (MFPT) approach. Similar to the thermodynamic
integration scheme, the MFPT approach utilizes the time-dependent
configurations as resulted from the independent runs under
identical conditions, however, the MFPT is focused on the averaged
time scales, at which a system characteristic (reaction
coordinate, order parameter) appears for the first
time~\cite{Mokshin/Barrat_JCP_2009,Mokshin/Barrat_PRE_2008,Wedekind_JCP_2007}.
We show that the thermodynamic integration scheme and the MFPT
method provide a convenient tool to treat the simulation results
(and/or the experimental data) concerning both the nucleation and
the growth kinetics. For the considered case of water, we define
the set of the characteristics for steady-state homogeneous
nucleation and growth of the liquid droplets on the basis of MD
simulation data.

\section{Numerical schemes}
\label{Num_schemes}

\textit{Thermodynamic integration.} -- The surface energy $w$ can
be defined as an excess energy per unit area of the surface that
is conditioned by the lack of neighbors for the surface particles
in comparison with the bulk particles (see
Ref.~\cite{J_Frenkel_book_1946}). If one restricts the
consideration by the closest neighbors only with the pairwise
additive interactions $u(r_{ij})$, then the following relation
appears directly
\begin{equation} \label{eq_energy_surf}
w = \frac{1}{2}u(\widehat{r}_{ij})(z - z')n',
\end{equation}
where $\widehat{r}_{ij}$ is the average distance between the
neighbors in a new phase, the quantity $n'$ denotes the number of
surface particles per unit area and depends on the size of a
nucleus, $z$ and $z'$ are the first coordination number of bulk
and surface particles, respectively. Then, the surface tension can
be estimated directly by the thermodynamic integration of the
surface energy as
\begin{equation} \label{surf_tens}
\sigma = - \int_{\lambda = 0}^{1} \left \langle \frac{\partial w}{
\partial \lambda} \right \rangle_{\lambda} d\lambda.
\end{equation}
The reaction coordinate $\lambda$ or the so-called
$\lambda$-scaling~\cite{Hansen/McDonald_book_2006} is associated
with the rescaled cluster size, $\lambda = (n/n^*)^{1/3}$, which
is equal to zero if there are no nuclei in the system and to unity
if the nucleus size has the critical value $n^*$. The notation
$\langle \cdots \rangle_{\lambda}$ means an ensemble average at a
particular value of $\lambda$.

\textit{Extended mean first passage time method.} -- According to
the continuous Zeldovich-Frenkel scheme, the nucleation process
can be described within a Fokker-Planck-type equation
\begin{equation}
\label{Fokker_Plank_eq} \frac{\partial N_n(t)}{\partial t} = -
\frac{\partial J_n}{\partial n} = \frac{\partial}{\partial n}
\left \{ N_n^{eq} g_n^+ \frac{\partial}{\partial n} \left [
\frac{N_n(t)}{N_n^{eq}} \right ] \right \},
\end{equation}
where $n$ is the cluster size, $N_n(t)$ is the time-dependent
cluster size distribution over unit volume, $J_n$ is the current
over cluster size space, $g_n^+$ is the monomer attachment rate to
a $n$-sized cluster and $N_n^{eq} = N_0^{eq} \exp(-\beta \Delta
G_n)$ is the equilibrium cluster size distribution,  $\Delta G_n$
is the work required to form the $n$-sized cluster and $\beta =
1/(k_B T)$.

If one considers the $n$-dependent term $\Delta G_n$, the
nucleation regime is directly associated with the vicinity of
critical value of the cluster size, $n^*$, where the term $\Delta
G_{n^*}$ corresponds to a nucleation barrier and has a maximum.
Assuming that the nucleation barrier can be expanded into the
Taylor series in this vicinity
\begin{equation}
\Delta G_n = \Delta G_{n^*} + \left .
\sum_{k=2}\frac{(n-n^*)^{k}}{k!}\frac{\partial^k \Delta
G_n}{\partial n^k} \right |_{n=n^*},
\end{equation}
the approximated evaluation of Eq.~(\ref{Fokker_Plank_eq}) in the
vicinity of nucleation regime can be written as
\begin{eqnarray} \label{MFPT_general}
J_{n'}^{-1} &=& \frac{\exp(\beta \Delta G_{n^*})}{g_{n^*}^+
N_0^{eq}
} \\
& & \times \int_0^{n'} dn \; \exp \left [\beta \left .
\sum_{k=2}\frac{(n-n^*)^{k}}{k!}\frac{\partial^k \Delta
G_n}{\partial n^k} \right |_{n=n^*}\right ]. \nonumber
\end{eqnarray}
The series in the exponential  of Eq.~(\ref{MFPT_general})
contains an information about the geometrical peculiarities of the
term $\Delta G_n$ around its maximum at $n^*$. Namely, the second
contribution of the series is related with the Zeldovich factor
$Z$ and characterizes the curvature of the barrier at the top
\begin{equation}
 - \frac{\beta}{2} \left . \frac{\partial^2 \Delta G_n}{\partial n^2} \right
|_{n=n^*} = \pi Z^2.
\end{equation}
Moreover, the ratio of the third and the second contributions,
which is $\Delta G^{(3)}_{n=n^*} / 3 \Delta G^{(2)}_{n=n^*}$,
indicates on the asymmetric properties of the barrier. For
example, if the ratio is equal to zero, then the barrier is
symmetric one and can be approximated by a parabolic geometry.
This means for the given example that we are restricted here only
by a case with $k=2$, which corresponds to the Zeldovich
approximation. Here, the analytical expression for the
steady-state nucleation rate $J_{s}$ can be directly obtained from
Eq.~(\ref{MFPT_general}) within the MFPT
method~\cite{Hanggi_RMP_1990}, where the averaged time scale of
the first appearance of the $n$-sized cluster $\tau_n^{MFPT}$ is
considered:
\begin{eqnarray} \label{MFPT_err_func}
\tau_n^{MFPT} &=& \frac{1}{2 J_{s} V}\{1+\mathrm{erf}[\sqrt{\pi}Z(n-n^*)]\}\\
&=& \frac{1}{2 J_{s} V}\mathrm{erfc}[\sqrt{\pi}Z(n-n^*)].
\nonumber
\end{eqnarray}
Here, $V$ is the system volume, and $\mathrm{erf}(x)=2
\pi^{-1/2}\int_{0}^{x}\exp(-t^2)dt$ is the error function.

\begin{figure}[ht!]
%\vspace*{0.5cm}
\begin{center}
% \begin{minipage}[c]{14cm}
    \includegraphics[width=9.5cm]{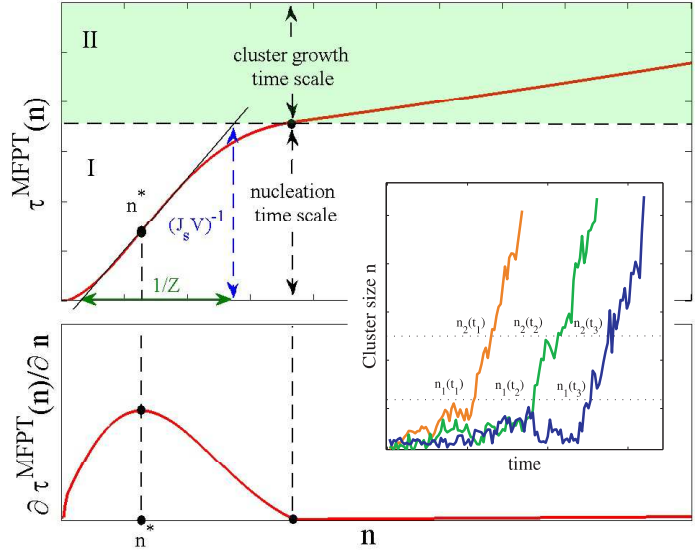}
%    \end{minipage}
%\vspace*{0.5cm}
%\newline
\end{center}
\caption{(Color online) Top: Schematic plot of the MFPT
distribution for the cluster size $n$ as obtained from simulation
(or experimental) data. The regions I and II  are associated with
nucleation and cluster-growth regimes, respectively. The routine
for finding the nucleation characteristics from the MFPT-curve is
presented. The gentle slope of the MFPT-curve at the transition
value, $n=n^*$, is evidence of the smooth form of the nucleation
barrier $\Delta G_n$ in the vicinity of $n^*$, while the location
of the inflection point (full circle) below the half-height
$1/(2J_sV)$ indicates qualitatively on the barrier asymmetry. The
pronounced increase of MFPT-curve in the region II appears due to
the fact that nucleation and cluster-growth time scales are
comparable, and this part of the curve as an inverted one,
$n(\tau^{MFPT})$, can be used to estimate the parameters of
cluster growth kinetics. Bottom: First derivative of the MFPT
distribution, $\partial \tau_n^{MFPT}/\partial n$. Here, the
maximum is associated with the inflection point, which is directly
located at the critical value of cluster size, $n=n^*$. Position
of the next extremum (minimum) on $(\partial
\tau_n^{MFPT}/\partial n)_{n>n^*}$ corresponds to the nucleation
time scale $\tau_n = 1/(2J_sV)$ as defined from the main MFPT
distribution $\tau_n^{MFPT}$. Inset: Typical cluster growth curves
obtained from the independent simulations.}\label{MFPT_scheme}
\end{figure}

The MFPT method provides the next useful capabilities in the
treatment of the nucleation-growth processes. The first one is
related with the critical value $n^*$, which is located at the
inflection point, i.e. at the point, where the first derivative
$(\partial \tau_n^{MFPT}/\partial n)_{n=n^*}$ has a maximum. Thus,
a simple analysis of $\partial \tau_n^{MFPT}/\partial n$ yields
the critical value $n^*$ (see Fig.~\ref{MFPT_scheme}). For the
particular case of Eq.~(\ref{MFPT_err_func}), one obtains directly
that $n=n^*$, when $\tau_{n=n^*}^{MFPT} = 1/(2 J_s V)$ that is the
consequence of the nucleation barrier symmetry. The second
property is that the Zeldovich factor can be directly extracted
from MFPT as
\begin{equation}
\label{Zeldovich_MFPT} Z = \left . J_s V \frac{\partial
\tau_n^{MFPT}}{\partial n}\right |_{n=n^*}.
\end{equation}
The geometric constructions, corresponding to this equation, are
presented in Fig.~\ref{MFPT_scheme}.
Equation~(\ref{Zeldovich_MFPT}) indicates that the smaller values
of $Z$ are resulted from the smaller values of $(\partial
\tau_n^{MFPT}/\partial n)_{n=n^*}$ at the fixed $J_s V$. On the
other hand, the smaller values of the Zeldovich factor correspond
to the flatter nucleation barrier curve $\Delta G_n$ near the
critical size $n^*$. And, finally, the third property is
associated with the steady-state nucleation rate $J_s$, which can
be defined from the MFPT distribution as $J_s = 1/(\tau_{n}^{MFPT}
V)$ at $n$, where $(\partial \tau_n^{MFPT}/\partial n)_{n>n^*}$
approaches the minimum and the distribution $\tau_{n}^{MFPT}$
starts itself to demonstrate a steady-like $n$-dependence (see
Fig.~\ref{MFPT_scheme}). Thus, using the known mean first passage
time distribution $\tau_n^{MFPT}$ one can directly define the
critical value $n^*$, the Zeldovich factor $Z$ and the
steady-state nucleation rate $J_s$ by a direct numerical analysis.

Nucleation-growth kinetics is characterized by the nucleation time
scale $\tau_{n} = 1/(J_s V)$ and the cluster-growth time scale
$\tau_{gr}$. The ratio between these time scales distinguishes the
separate cases for the numerical treatment within the MFPT method:
(i) If $\tau_n \gg \tau_{gr}$, then $\tau_{n}^{MFPT}$ demonstrates
a clear defined plateau of the height $\tau_{n} = 1/(J_s V)$, that
simplifies significantly accurate estimation of the nucleation
rate; (ii) If these time scales are comparable, $\tau_n \sim
\tau_{gr}$, then the errors can appear in the estimation of
$\tau_{n}$, since the boundary between nucleation and growth in
MFPT distribution is smeared.

Furthermore, the MFPT method gives a convenient tool to extract
the characteristics of nucleus growth kinetics, which follows the
nucleation regime in the MFPT distribution (see
Fig.~\ref{MFPT_scheme}). In fact, the inverted MFPT distribution,
$n(\tau^{MFPT})$, has the statistical meaning of the most probable
cluster growth law for the growth regime of the MFPT curve.

Following Ref.~\cite{Kashchiev_Nucleation}, the growth law of a
cluster can be taken in general form as
\begin{equation}
R(t) = R_* + (\mathcal{G}_{cl} t)^{\nu},
\end{equation}
where $R$ and $R_*$ is the radius of the growing cluster and the
critically-sized cluster, respectively;   $\nu$ is the growth
exponent and $\mathcal{G}_{cl}$ is the growth constant, which has
a dimension of $[m^{1/\nu}/s]$. Then, the growth rate is $G(t) =
\nu \mathcal{G}_{cl}^{\nu}t^{\nu-1}$, while the acceleration of a
cluster growth can be formally defined as $a(t) = \nu (\nu -1)
\mathcal{G}_{cl}^{\nu}t^{\nu-2}$. The steady cluster growth with a
constant growth rate corresponds to the particular case of
$\nu=1$, where the growth rate coincides with the growth factor,
i.e. $G(t) = \mathcal{G}_{cl} = \mathrm{const}$, otherwise (at
$\nu \neq 1$) one has the process with unsteady growth
rate~\cite{Mokshin/Barrat_PRE_2010}. Further, taking into account
that the volume of a growing cluster evolves with time  as $V(t) =
c_g [R(t)]^3$ and $N(t) = \rho_{cl} V(t)$, where $c_g$ is a
dimensionless cluster-shape factor ($c_g = 4\pi/3$ in a case of
the sphere) and $\rho_{cl}$ is the density of the cluster-phase,
one can write the growth law in the extended form:
\begin{equation} \label{eq_growth_law_ex}
n(t,t_c) = n^* \left [1 +
\mathcal{G}_{cl}^{3\nu}(t-t_{c})^{3\nu}\frac{\rho_{cl}c_g}{n^*} +
3\mathcal{G}_{cl}^{2\nu}(t-t_{c})^{2\nu}\left (
\frac{\rho_{cl}c_g}{n^*} \right )^{\frac{2}{3}} +
3\mathcal{G}_{cl}^{\nu}(t-t_{c})^{\nu} \left (
\frac{\rho_{cl}c_g}{n^*} \right )^{\frac{1}{3}} \right ].
\end{equation}
Here, the lag-time $t_c$ defines the appearance of the
critically-sized cluster. Then, the term $n(\tau^{MFPT})$ can be
fitted for the growth regime by Eq.~(\ref{eq_growth_law_ex}) to
extract the growth characteristics: the cluster-shape factor
$c_g$, the growth constant $\mathcal{G}_{cl}$ and the growth
exponent $\nu$. At rapid growth of small clusters the last two
contributions in Eq.~(\ref{eq_growth_law_ex}) can be neglected,
and the growth law takes the form~\cite{Mokshin/Barrat_PRE_2010}
\begin{equation} \label{eq_growth_law}
n(t,t_c) \simeq n^* + c_g \rho_{cl}\mathcal{G}_{cl}^{3\nu}(t -
t_c)^{3\nu},
\end{equation}
where $\nu$ is positive.
\begin{figure}[ht!]
%\vspace*{0.5cm}
\begin{center}
% \begin{minipage}[c]{12cm}
    \includegraphics[width=9.5cm]{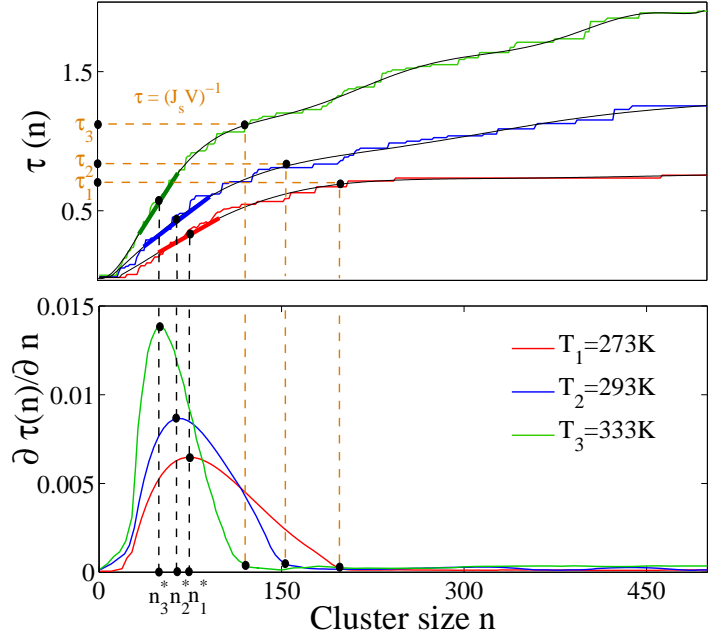}
%    \end{minipage}
%\vspace*{0.5cm}
%\newline
\end{center}
\caption{(Color online) Top: Direct MFPT distributions
(stepwise-curves) and their interpolations (smooth solid curves)
for the temperatures $T=273$, $293$ and $333$\;K. Dots on the
curves indicate the inflection points, which define the critical
sizes $n^*$, and the time scales $\tau=(J_s V)^{-1}$ corresponding
to the nucleation rates. Thick short lines are the linear parts of
the interpolated curves near $n^*$ and define the ranges of errors
in critical sizes. Note that the errors in the nucleation rates
can be also defined as a result of changes in $\tau=(J_s V)^{-1}$
due to the correction of different interpolations with the same
accuracy in the reproduction of the direct MFPT-distributions.
Bottom: First derivative of the MFPT distributions, $\partial
\tau_n^{MFPT}/\partial n$, for the same temperatures.
}\label{MFPT_part}
\end{figure}

\section{Computational details}

Molecular dynamics simulations were performed in the spirit of
previous studies of the structural transformations in this system
described in Refs.~\cite{Moore_Nature_2011,Molinero_JPCB_2009}
with only difference in the details related with the considered
thermodynamic range. We have examined the system composed
$N=8\;000$\;particles (molecules) interacting via the mW-potential
in the cubic cell with the periodic boundary conditions in all
directions. The time-step for numerical integration was $1$\;fs;
and the $NpT$ (number, pressure, temperature) ensemble was applied
with $p=1$\;atm. Pressure and temperature were controlled via the
Nos\'{e}-Hoover barostat and thermostat, respectively, acting
uniformly throughout the system. The damping thermostat and
barostat constants were taken to be $\tau_T = \tau_P = 10$\;fs.
The parameters of the mW-potential are completely identical to
those reported in
Refs.~\cite{Moore_Nature_2011,Molinero_JPCB_2009}.

Initially, the set of a hundred of independent samples was
prepared and equilibrated at the temperature $T=900$\;K on the
time scale $50$\;ps (i.e. $50\;000$ time-steps). The
correspondence of the systems to the vapor phase was directly
confirmed by the particle diffusivity and the distinctive particle
radial distribution functions. Moreover, following
Ref.~\cite{Moore_Nature_2011}, the samples were cooled at
$10$\;K/ns to the desired temperatures from the range $273 \leq
T/K \leq 373$ (at $p \simeq 1$\;atm.).\footnote{It is necessary to
note that the mW-model reproduces correctly equation of state
$\rho(T)$ for this temperature range (see Fig.~$4$ in
Ref.~\cite{Molinero_JPCB_2009}).} Then, over a time scale $\sim 1
\div 10$\;ps each a system was `equilibrated' till the
disappearance of the pronounced fluctuations in temperature and
pressure, after that the initial configurations were stored for
the further study of the vapor-to-liquid nucleation process. Note
that this cooling procedure is similar to the reported one in
Ref.~\cite{Matsubara_JCP_2007}. The following  $NpT$-simulations
starting from these configurations -- a hundred for each
considered temperature -- were performed to collect the statistics
of the independent nucleated events, where the time-dependent
cluster size distributions $N_n(t)$ were evaluated (for an every
run). The averaged time scale for the simulations in this
nucleation-growth regime was $50$\;ns. On the basis of the found
$N_n(t)$-distributions, the MFPT-curves were extracted and the
nucleation characteristics were estimated according to the scheme
presented above. After this, the critical sizes $n^*$ defined from
the MFPT-curves were used at the retreatment of the simulation
data with the aim to define the distributions of the energy
$\omega$ over the reaction coordinate $\lambda$.

An identification of the particles, which belong to liquid phase,
was performed in the spirit of the Stillinger
rule~\cite{Stillinger_JCP_1963}. First, the particles are
``\textit{neighbors}'' (or bonded) if the distance between their
centers is less than $r_s$, where $r_s$ is the position of the
first minimum in the pair correlation function of the liquid phase
(at the same conditions). Further, a particle is considered as a
liquid-like if it has, at least, four neighbors~\footnote{The last
condition allows one to remove from the consideration those
particle-pairs, which are result of the instant random event and
are not related to the formation of a new phase.}.
\begin{figure}[ht!]
%\vspace*{0.5cm}
\begin{center}
% \begin{minipage}[c]{13.7cm}
    \includegraphics[width=9.9cm]{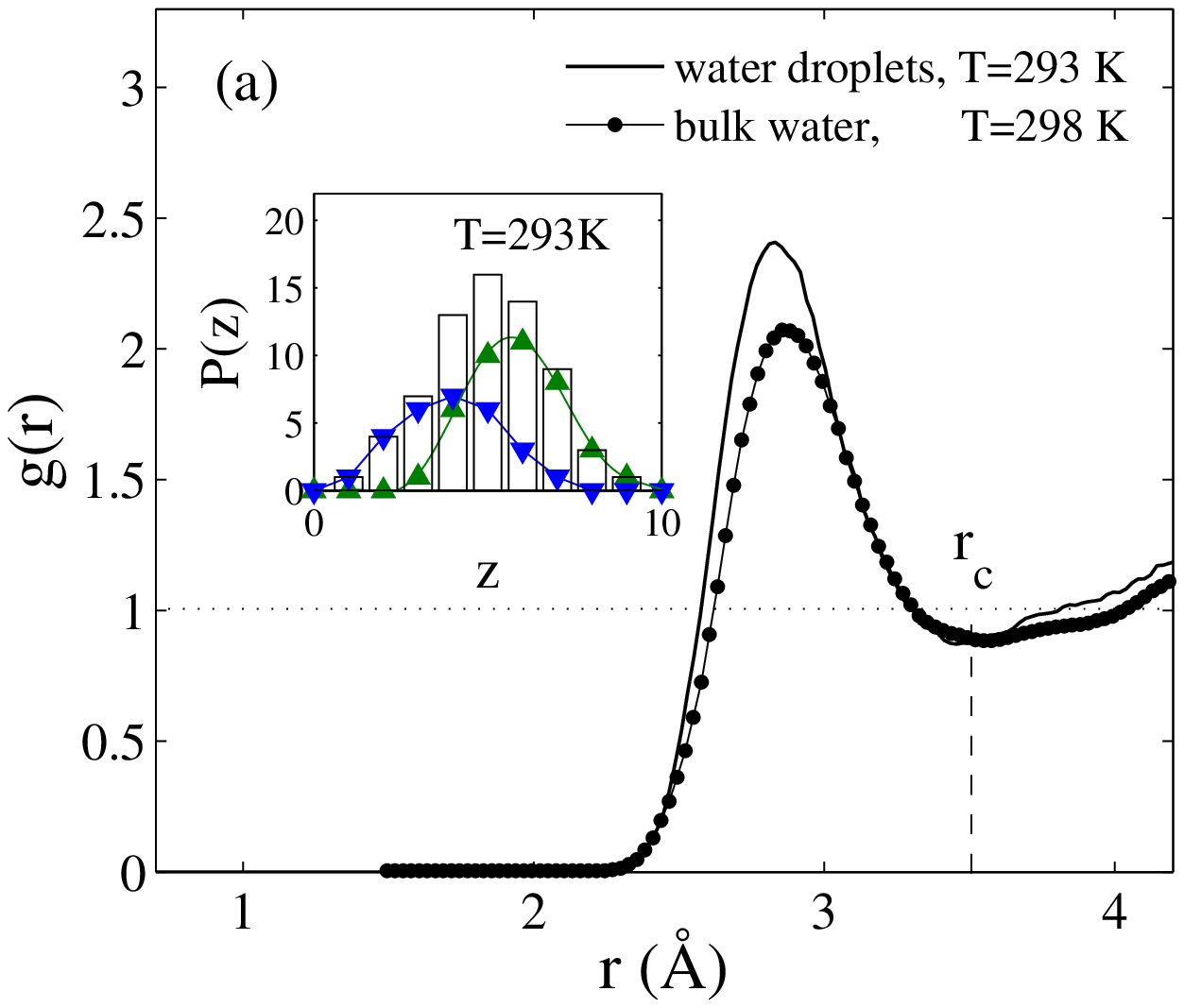}
    \includegraphics[width=9.9cm]{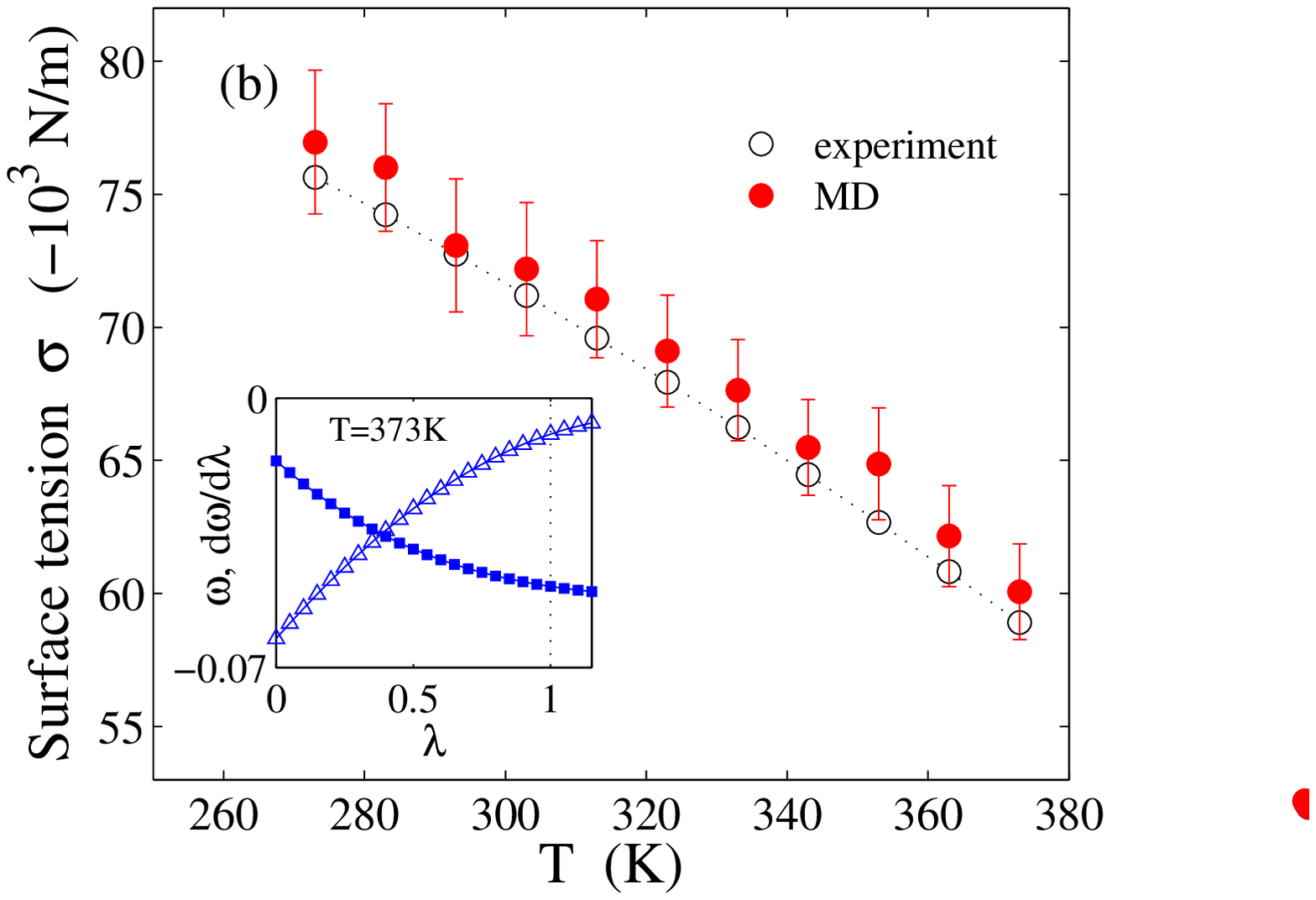}
\end{center}
\caption{\label{fig_surf_tension} (Color online) (a) Radial
distribution function $g(r)$ of liquid water as resulted from the
mW-model: results for bulk water at $T=298$\;K and $p=0$ (full
circles) reported in Ref.~\cite{Molinero_JPCB_2009}, data for the
bulk range of critically-sized droplets at $T=293$\;K and
$p=1$\;atm. Inset: Distribution of the first coordination number
for the water molecules of a critically-sized liquid droplet at
the temperature $T=293$~K. The histogram corresponds to the total
distribution; the line with triangles present an impact from the
bulk molecules, $z$; and the line with rotated triangles shows the
contribution of the surface molecules, $z'$. The data are averaged
over set of runs.  (b) Main: Temperature dependence of the surface
tension $\sigma$. The simulation results show the averages (full
circles) and standard deviations (error bars) from independent
runs; experimental data are presented by open circles, whereas the
dotted line is the interpolation by $\sigma(T) = B[(T_c -
T)/T_c]^m \{ 1+b[(T_c-T)/T_c] \}$, $B=235.6$~N/m, $b=-0.625$,
$m=1.256$ and $T_c=647.15$~K \cite{Surf_tension_textbook}. Inset:
Surface energy and slope of the surface energy $\partial
\omega/\partial \lambda$ as functions of $\lambda= (N/N_c)^{1/3}$
for a growing water droplet. At the critical size $N=N_c$ one has
$\lambda = 1$. The presented results are outcome of a single
simulation run.}
\end{figure}

\section{Results} \label{sec_results}

In Fig.~\ref{MFPT_part} the obtained MFPT distributions
$\tau_n^{MFPT}$ and their derivatives $\partial
\tau_n^{MFPT}/\partial n$ at the temperatures $T=273$, $293$ and
$333$\;K are presented as an example. On the basis of the defined
values of the critical size $n^*$ and nucleation rate $J_s =
(\tau^{MFPT} V)^{-1}$, the values of the Zeldovich factor were
extracted within Eq.~(\ref{Zeldovich_MFPT}). The typical
$\lambda$-dependences of the surface energy and its derivative as
obtained from a single run are shown in the inset of
Fig.~\ref{fig_surf_tension}(b). The averages of the slope $\langle
d \omega/ d \lambda \rangle_{\lambda}$ over different runs were
used to estimate Eq.~(\ref{surf_tens}) by means of the trapezoidal
method. The smooth character of the curves allows one to restrict
oneself by the method and to exclude higher order integration
schemes~\cite{Ytreberg_JCP_2006}. It is necessary to note that
errors in the critical size $n^*$  have not been considered at the
estimation of the surface tension with Eq.~(\ref{surf_tens}).

\textit{Coordination number.} -- The inset of
Fig.~\ref{fig_surf_tension}(a) shows the distribution of the first
coordination number for the water molecules generated a droplet of
the critical size in the system at the temperature $T = 293$~K.
The presented histogram is the cumulative result of the
distributions for surface and bulk molecules. Remarkably, these
distributions (for bulk and surface molecules) are symmetric ones
as well as reproducible by the Gaussian functions. Further, the
averaged values of the coordination numbers $z$ and $z'$ are
extracted from the distributions and appear to be $5.8$ and
$3.93$, respectively, for the case. Other important observation is
that the term $z$ is practically unchangeable with temperature,
whereas the coordination number in a surface layer demonstrates a
smooth insignificant decrease with the temperature increasing [for
comparison, from $z'(T=293\;\mathrm{K}) = 3.93$ to
$z'(T=353\;\mathrm{K})=3.47$].

Moreover, the value of the coordination number for bulk molecules
in the droplets coincides with the found value of $z$ extracted on
the basis of the integral
definition~\cite{Khusnutdinoff/Mokshin_PhysicaA_2012}
\begin{equation}
z = 4 \pi \rho_l \int_0^{r_c} r^2 g(r) dr,
\end{equation}
where $r_c$ is the first minimum position in the radial
distribution function $g(r)$. Nevertheless, the found value
$z(T=293\;\mathrm{K})=5.8$ differ from the result of Molinero and
Moore~\cite{Molinero_JPCB_2009} obtained within the mW-model for
the bulk water, $z(T=298\;\mathrm{K})=5.1$. To understand the
reasons of the discrepancy, we compare the corresponding radial
distribution functions in Fig.~\ref{fig_surf_tension}(a). As can
be seen, the intensity of the first maximum of $g(r)$ is higher
for the case of the water droplets, although the maximum is
located at a lower distance. This feature indicates that the
liquid phase is characterized by the more pronounced short-range
ordering for the case of the microscopically \emph{small
nucleated} clusters than for the equilibrium liquid phase
considered in Ref.~\cite{Molinero_JPCB_2009}.

\textit{Surface tension.} -- In Fig.~\ref{fig_surf_tension}(b),
the temperature dependence of the surface tension of the
critically-sized nuclei is presented. The results obtained from
simulation data demonstrate the known decrease of the surface
tension with temperature~\footnote{According to results of
Ref.~\cite{Molinero_JPCB_2009} obtained for the planar surface
tension at a single temperature $T=300$~K, the mW-model gives the
best agreement with the experiment in comparison to the models:
SPC, SPC/E, TIP$3$P, TIP$4$P, TIP$5$P.} and reproduce precisely
the experimental data of this term for a planar liquid-vapor
interface \cite{Surf_tension_textbook}. Such a good agreement of
our results with experimental data is unexpected because of the
two next reasons, mainly. First, in contrast to an inherent water
system, the mW-model excludes the long-range intermolecular
interactions, which still can have an influence on the interface
effects. Second, no adjustment of simulation data was performed to
take into account the finite size effects. Thereby, the
corrections to surface free energy in the spirit of the Tolman's
\emph{ansatz}, which are appeared to be proportional to the
inverse linear size of the critical nucleus, were unconsidered for
the surface tension.

On the other hand, the surface tension of a planar interface was
recently defined with the mW-model for the temperatures $250 \leq
T/K \leq 350$~\cite{Molinero/Baron}. The values of the surface
tension reported in Ref.~\cite{Molinero/Baron} have the lower
values in comparison with the values presented in
Fig.~\ref{fig_surf_tension}(b) for the water droplets, and the
difference is about $5 \div 10$\;percents for the same temperature
range. Although this difference could be attributed to the Tolman
length with the negative value, like it was reported by Kiselev
and Ely~\cite{Kiselev/Ely_PhysicaA_2001} for the liquid-ice
surface tension, we assume that our values of $\sigma$ can be
overestimated because of the neglect the three-particle
interactions by computational protocol within
Eq.~(\ref{eq_energy_surf}).

Nevertheless, the surface tension decrease with the temperature is
directly consistent with the temperature decreasing of
$u(\widehat{r}_{ij})$, of the surface coordination number [or the
increase of the difference $(z-z')$] and of the surface particle
density, where the last two contributions are practically
counterbalanced by each other~\footnote{Assuming the spherical
droplet of the radius $R$ with the thickness of the surface layer
$\Delta$, the surface particle density can be roughly estimated
$n'(R) \propto [R - (R-\Delta)^3/R^2$].}.

Recalling the previous debates \cite{Lu_APL_2006}, the temperature
range investigated here can contain the inflection points in the
vapor-liquid surface tension of water. The results, presented in
Fig.~\ref{fig_surf_tension}(b), indicate on the absence of the
clear detected inflections in the $T$-dependence of the surface
tension.

\begin{figure}[ht!]
%\vspace*{0.0cm}
\begin{center}
    \includegraphics[width=12.0cm]{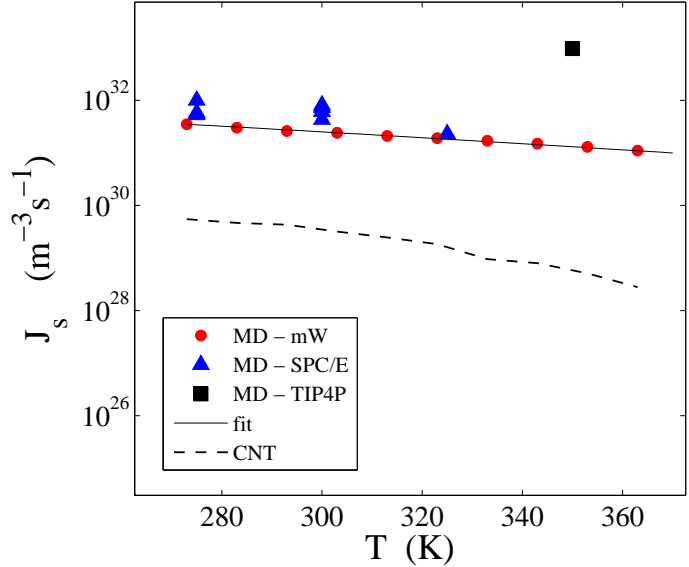}
%\vspace*{-0.5cm}
%\newline
\end{center}
\caption{\label{fig_nucl_rate} (Color online) Temperature
dependence of the homogeneous droplet (vapor-to-liquid) nucleation
rate $J_s(p\simeq1\;\mathrm{atm},T)$ in water. Comparison of the
simulation results with the mW-model [the vapor density is
$\rho_{v} \in [1.12;\; 1.55]\times 10^{-2}$nm$^{-3}$], with the
atomistic SPC/E-model from Ref.~\cite{Matsubara_JCP_2007} [the
density is $\rho_{v} \in [1.23;\; 1.86]\times 10^{-2}$nm$^{-3}$],
with the atomistic TIP$4$P-model from
Ref.~\cite{Yasuoka/Matsumoto_JCP_1993} [the numerical density is
$\rho_{v} = 1.55 \times 10^{-2}$nm$^{-3}$] and the treatment of
nucleation data within the CNT. Solid line show the best fit by
means of the function $J_s(p \simeq 1\;\mathrm{atm},T) \propto
\exp(-0.00092 \; T^{1.4})$.}
\end{figure}
\textit{Steady-state nucleation rate.} -- In
Fig.~\ref{fig_nucl_rate}, the steady-state nucleation rates
obtained from the MFPT treatment of the simulation data with the
mW-model are presented as a function of temperature (see also
Table~\ref{Tab_1}). The presented data cover the density range
$\rho_{v}$ from $1.14 \times 10^{-2}$nm$^{-3}$ to $1.55 \times
10^{-2}$nm$^{-3}$. As can be seen, the values of $J_s(T)$ compare
well with those obtained for the same density range by Matsubara
\textit{et al.} with the atomistic
SPC/E-model~\cite{Matsubara_JCP_2007}. Nevertheless, as contrasted
to results of Ref.~\cite{Matsubara_JCP_2007}, which are scattered
over ($J_s,T$)-plot, the values of the nucleation rate $J_s$
obtained within the mW-model demonstrate the smooth decrease with
temperature over the considered temperature range. Moreover, this
decrease is well-reproduced by the dependence $\ln[J_s(T)]= -
0.00092\;T^{1.4} +75$.  Among all the data presented on
Fig.~\ref{fig_nucl_rate}(a), the highest value of the nucleation
rate appears from the simulations of
Ref.~\cite{Yasuoka/Matsumoto_JCP_1993} with the TIP$4$P-model.

On the other hand, it is attractive to test the dependence
$J_s(T)$ within the CNT treatment with the extracted values of the
other nucleation characteristics. So, the original Becker-D\"oring
formulation yields
\begin{equation} \label{eq_nucl_rate_BD}
J_{s}^{CNT} = \frac{\rho_v^2}{\rho_l}
\sqrt{\frac{2\sigma_{\infty}}{\pi m}} \exp \left ( - \frac{\Delta
G_{n^*}}{k_B T} \right ),
\end{equation}
where the barrier can be taken as
\begin{equation} \label{eq_barrier}
\frac{\Delta G_{n^*}}{k_B T} = 3 \pi (n^* Z)^2,
\end{equation}
and $\sigma_{\infty}$ is the surface tension for a planar
liquid-vapor interface~\cite{Surf_tension_textbook}. As can be
seen from Fig.~\ref{fig_nucl_rate}, although the both dependencies
demonstrate a similar behavior decaying with temperature, the pure
simulation results for nucleation rate have in two orders higher
values in comparison with $J_{s}^{CNT}$. This is evidence of the
difficulties at the description of the droplet nucleation in water
by means of the CNT, which are similar with those reported earlier
for the studies with the atomistic
models~\cite{Chen_JCPA_2005,Matsubara_JCP_2007} as well as with
the experimental data (see Fig.3 of Ref.\cite{Hale_JCP_2005}).

\begin{figure}[ht!]
%\vspace*{0.5cm}
\begin{center}
% \begin{minipage}[c]{14.8cm}
    \includegraphics[width=12cm]{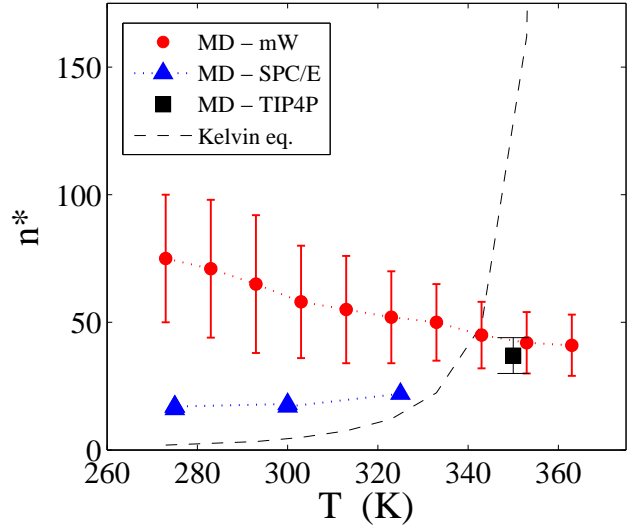}
%    \end{minipage}
%\vspace*{0.5cm}
%\newline
\end{center}
\caption{\label{fig_cluster_size} (Color online) Temperature
dependence of the critical cluster size, determined from
simulations within the different models of potential fields
(mW-model, SPC/E~\cite{Matsubara_JCP_2007},
TIP$4$P~\cite{Yasuoka/Matsumoto_JCP_1993}), compared to prediction
by Kelvin equation. In the case of the mW-model, error bars are
defined by a width of the curvature range in MFPT-distributions.}
\end{figure}
\textit{Critical cluster (droplet) size.} --
Figure~\ref{fig_cluster_size} illustrates the temperature
dependence of the critical cluster size $n^*$, where the
simulation results with the mW-model are compared with the
simulation data of Matsubara \textit{et al.}
\cite{Matsubara_JCP_2007} and of Yasuoka \textit{et al.}
\cite{Yasuoka/Matsumoto_JCP_1993} as well as with the predictions
of the Kelvin equation
\begin{equation} \label{Kelvin_equation}
n^* = \frac{32 \pi}{3} \frac{\sigma_{\infty}^3}{\rho_l^2 [k_B T
\ln (p/p^s)]^3}.
\end{equation}
Here $p^s$ is the saturated water vapor
pressure~\cite{Alexandrov_book_1999}.

\begin{figure}[ht!]
%\vspace*{0.5cm}
\begin{center}
% \begin{minipage}[c]{16cm}
    \includegraphics[angle=90,width=9.5cm]{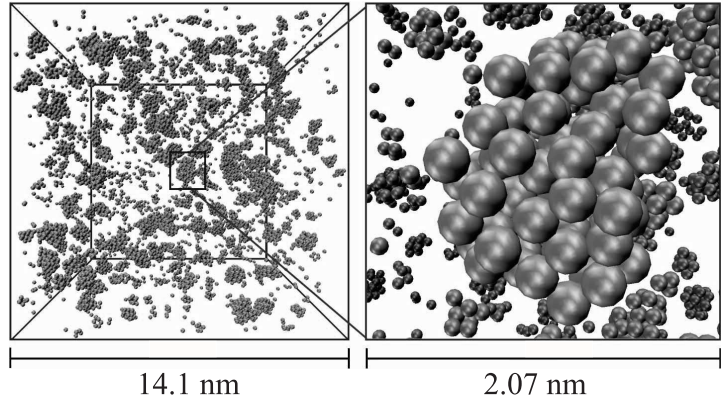}
% \end{minipage}
%\vspace*{0.5cm}
%\newline
\end{center}
\caption{\label{fig_snapshot} (Color online) Left: Snapshot of the
water system at the temperature $T=293$~K and the pressure
$p=1$~atm. at the moment, when the critically-sized droplets are
appearing. Right: Growing droplet of the same system.}
\end{figure}

First, for the mW-model the critical cluster size reveals a slight
decrease with temperature from $n^* = 75$ to $40$ particles over
the temperature range $273 \leq T/K \leq 363$. This change of the
cluster size means the decrease of the droplet radius from $4$ to
$3.3$ of the averaged water molecule diameters. Obviously, the
change is insignificant. Moreover, the observed decrease is masked
by errors, which were defined as the curvature range width in the
MFPT-distributions (see Fig.~\ref{MFPT_part}). The range of errors
is $ \pm(10 \div 25)$ particles, that is awaited to be reasonable,
since it covers only a few surface part of a water droplet
(Fig.~\ref{fig_snapshot}). In addition, according to the
definition within the MFPT-method these errors should be
considered as the probable deviations from $n^*$ in a statistical
sense. The comparison with the results obtained for the atomistic
models (TIP$4$P and SPC/E) reveals that the values of $n^*$
obtained within the mW-model overestimate the data of the
SPC/E-model, but are in agreement with a single value found by
Yasuoka \textit{et al.} in simulations with the TIP$4$P-model.

Further, the mW-model result for $n^*(T)$-curve is different from
the predictions of Eq.~(\ref{Kelvin_equation}), which yields the
increase of $n^*$ with the temperature $T$ (see
Fig.~\ref{fig_cluster_size})~\footnote{The evaluation of the
supersaturation $S=p_v/p_v^s$ was performed with the experimental
values of the saturated water vapor pressure $p_v^s$.
Nevertheless, one needs to note that the mW-model can yields the
results different from the experimental data for $p_v^s(T)$.}.
However, it should be noted that as far as the predictions of the
Kelvin equation are concerned,  it gives the values $n\simeq 2
\div 12$ molecules for the temperature range $273 \leq T/K \leq
363$ and the pressure $p = 1$~atm. In terms of the linear cluster
sizes, these values correspond to $1\div2$ water molecule
diameters. It is clear that the treatment of the stability of such
a small cluster from the thermodynamic point of view, which
requires the availability of the separated surface and bulk
regions of the cluster, is impossible.~\footnote{ We note that the
direct comparison of the predictions of the Kelvin equation with
simulation results should be considered as very approximate, since
the saturation curve resulted from a considered model can be
different from the real water saturation
curve.~\cite{Chen_JCPA_2005} For the mW-model, the additional
studies are necessary to clarify this point.}

\begin{figure}[ht!]
%\vspace*{2.0cm}
\begin{center}
% \begin{minipage}[c]{14cm}
    \includegraphics[width=12.0cm]{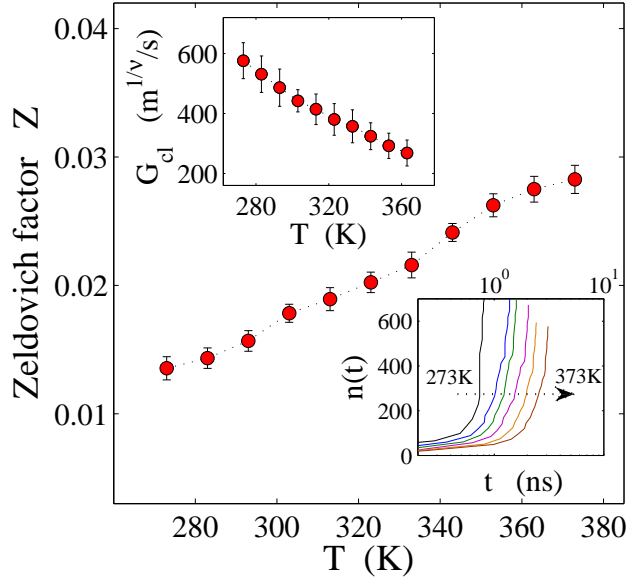}
%    \end{minipage}
%\vspace*{0.5cm}
%\newline
\end{center}
\caption{\label{fig_Zeldovich_factor} (Color online) Main:
Temperature dependence of the Zeldovich factor $Z$ as defined by
the MFPT method on the basis of the simulations with the mW-model.
Top inset: Temperature dependence of the growth factor
$\mathcal{G}_{cl}$, which is found from the fit of
Eq.~(\ref{eq_growth_law}) to the simulation data. The growth
exponent $\nu = 1.3$ and the term $A =
c_g\rho_l(\mathcal{G}_{cl}t_c)^{3\nu}/n^* = 1.16 \pm 0.2$ appear
 to be invariant respective the temperature. Error bars show the
standard deviations from the averages. Bottom inset: Growth curves
of the liquid water droplets emerging in the vapor phase at the
temperatures $273 \leq T/K \leq 373$.}
\end{figure}

\textit{Nucleation barrier and the Zeldovich factor.} -- The
temperature dependence of the next nucleation characteristic, the
Zeldovich factor $Z$, is presented in
Fig.~\ref{fig_Zeldovich_factor}. As can be seen, this quantity
decreases from the value $0.028$ to $0.014$ with the decrease of
the temperature $T$ (see also Table~\ref{Tab_1}). Such a behavior
is a direct evidence that the nucleation barrier loses its
sharpness and becomes more smoother with the decreasing the
temperature. This is qualitatively in an agreement with the
prediction of the CNT~\cite{Kashchiev_Nucleation}. Moreover, if
one assumes that the CNT yields the correct results for the
vapor-to-liquid nucleation of water then it is possible to define
the nucleation barrier by means of the simple
relation~(\ref{eq_barrier}). The direct evaluation yields the
correct tendency of the temperature dependence for the nucleation
barrier, though this tendency is not so pronounced one as it could
be expected for a sufficiently wide temperature range considered
here. According to the CNT, the nucleation driving force $|\Delta
\mu|$, growing with the decrease of $T$, must reduce the
nucleation barrier~\cite{Kashchiev_Nucleation}.

Within Eq.~(\ref{eq_barrier}) and simulation results one has that
the nucleation barrier decreases from $\beta \Delta G_{n^*}^{MD} =
12 \pm 2.7$ to $\beta \Delta G_{n^*}^{MD} = 9.7 \pm 2.5$ with the
decrease of the temperature from $T=363$ to $273$\;K. On the other
hand, the temperature dependence of the nucleation barrier as
predicted by the CNT is defined by
\[
\Delta G_{n^*} \propto \frac{\sigma_{\infty}^3}{\rho_l^2 |\Delta
\mu|^2},
\]
where $|\Delta \mu|$ is the chemical potential difference of
particles in the vapor and in the liquid phase. So, the observed
behavior of the nucleation barrier can be explained for the case,
where the change of $\sigma_{\infty}^3/\rho_l^2$ with the
temperature is completely counterbalanced by the change of
$|\Delta \mu|^3$.

Remarkably, the comparable values for the nucleation barrier arise
with the atomistic SPC/E model (see Table I in
Ref.~\cite{Matsubara_JCP_2007}), where the barrier changes from
$\beta \Delta G_{n^*} = 8.1$ to $\beta \Delta G_{n^*} = 6.7$ with
the temperature decreasing from $T=325$ to $275$\;K. Thus, the
observed results for the nucleation barrier can not be considered
as a consequence of the coarse-grained character of particle
interactions in the mW-model.
\begin{table}
  \caption{ Simulation results: system temperature $T$ (K); vapor
number density $\rho_{v}$ ($\times 10^{-2}$~nm$^{-3}$); critical
cluster size $n^{*}$; nucleation barrier $\Delta G/k_{B}T$;
nucleation rate $J_{s}$ ($\times 10^{32}$~m$^{-3}$s$^{-1})$; the
Zeldovich factor $Z$.}
   \label{Tab_1} \vspace*{0.5cm}
    \begin{center}
  \begin{tabular}{cccccc}
    \hline \hline
 $T$ & $\rho_{v}$ & $n_{c}$ & $\Delta G/k_{B}T$ & $J_{s}$ & $Z$\\
    \hline
273 & $1.548\pm0.095$ & $75\pm25$ & $9.72\pm 2.55$ & $0.35$ &  $0.0135 \pm 0.0009$ \\
283 & $1.462\pm0.060$ & $71\pm27$ & $9.76\pm 2.06$ & $0.30$ &  $0.0143 \pm 0.0008$ \\
293 & $1.426\pm0.077$ & $65\pm27$ & $9.78\pm 2.39$ & $0.26$ &  $0.0157 \pm 0.0008$ \\
303 & $1.425\pm0.051$ & $58\pm22$ & $10.07\pm 1.57$ & $0.24$ & $0.0182 \pm 0.0007$ \\
313 & $1.353\pm0.062$ & $55\pm21$ & $10.21\pm 2.25$ & $0.21$ & $0.0189 \pm 0.0009$ \\
323 & $1.313\pm0.047$ & $52\pm18$ & $10.43\pm 2.18$ & $0.19$ & $0.0202 \pm 0.0008$ \\
333 & $1.250\pm0.037$ & $50\pm15$ & $10.98\pm 2.53$ & $0.17$ & $0.0216 \pm 0.001$ \\
343 & $1.224\pm0.032$ & $45\pm13$ & $11.10\pm 1.70$ & $0.15$ & $0.0241 \pm 0.0007$ \\
353 & $1.174\pm0.045$ & $42\pm12$ & $11.50\pm 1.94$ & $0.13$ & $0.0262 \pm 0.0009$ \\
363 & $1.140\pm0.033$ & $41\pm12$ & $11.97\pm 2.17$ & $0.11$ & $0.0275 \pm 0.001$ \\
\hline \hline
\end{tabular}
\end{center}
\end{table}

\textit{Growth laws of the nucleated droplets.} -- Bottom inset of
Fig.~\ref{fig_Zeldovich_factor} shows the growth curves of the
liquid droplet at the different temperatures, which were found
from the statistical treatment of simulation data by means of the
MFPT approach as it was discussed above. Hence, these curves
depict the most probable growth laws in a statistical sense. As
can be seen from the figure, at the lower temperature the droplet
growth occurs faster. At the same time, all the curves are well
reproduced by Eq.~(\ref{eq_growth_law}), and the fitting to the
simulation data yields the following features. The growth exponent
$\nu$ in Eq.~(\ref{eq_growth_law}) appears to be invariant over
the temperature $T$ and it takes the value $\nu = 1.3$ at all the
considered temperatures. This indicates that the increasing the
linear size, i.e. the radius, averaged over all the directions of
the liquid droplet follows for all the temperatures the growth law
$R(t) = (\mathcal{G}_{cl}t)^{1.3}$, herewith, the droplet growth
itself is unsteady.

Further, the growth factor $\mathcal{G}_{cl}$ decreases with the
increase of the temperature $T$ (see top inset of
Fig.~\ref{fig_Zeldovich_factor}), that characterizes the faster
droplet growth at the lower temperatures. Following
Ref.~\cite{Mokshin/Barrat_PRE_2010}, Eq.~(\ref{eq_growth_law}) can
be written in the rescaled form:
\begin{equation}
\label{eq_rescaled_growth} n(\xi)/n^* \simeq 1 + A (\xi -
1)^{3\nu},
\end{equation}
where $A = c_g \rho_l (\mathcal{G}_{cl} t_c)^{3\nu}/n^*$ and $\xi
= t/t_c$ is the rescaled time. We found that in accordance with
the rescaled form~(\ref{eq_rescaled_growth}) all the growth curves
collapse onto a single curve independently of the temperature. In
addition, the parameter $A$ in Eq.~(\ref{eq_rescaled_growth})
takes the same value for all the considered temperatures, i.e.
$A=1.16 \pm 0.2$. This can be evidence of the generic features of
the water droplet growth process. Remarkably, this result is
correlated with the features of the crystal growth kinetics for a
model glassy system under shear drive, which were reported in
Ref.~\cite{Mokshin/Barrat_PRE_2010}.

\textit{Shape and sphericity of the nucleated droplets.} -- Other
issue, which is crucial in the CNT, is related with the shape and
the anisotropy of the growing
droplets~\cite{Reinhardt_JCP_2012,Tanguy_Review_2011}. A
convenient way to perform this study in our case is to use the
asphericity parameter in the next definition:
\begin{equation}
S_0 = \left \langle
\frac{(I_{xx}-I_{yy})^2+(I_{xx}-I_{zz})^2+(I_{yy}-I_{zz})^2}{2(I_{xx}+I_{yy}+I_{zz})^2}
\right \rangle, \nonumber
\end{equation}
where
\[
I_{\alpha\beta} = \sum_{i=1}^{n^*}m_0(r_i^2 \delta_{\alpha\beta} -
r_{i\alpha}r_{i\beta})
\]
is the components of the moment of inertia tensor associated with
a droplet, $m_0$ is the molecule mass, $\alpha,\beta \in
\{x,y,z\}$ are the components of the vector $\vec{r}$ between the
droplet center-of-mass and molecule $i$; the brackets $\langle
\ldots \rangle$ mean the statistical average over critically-sized
droplets of the different simulation runs. This definition of a
asphericity parameter sets the characterization: for a spherical
droplet one has $S_0 = 0$, whereas for an elongated and
string-like cluster one obtains $S_0 \to 1$. We found that
independently on the particular conditions (temperature, vapor
density) the asphericity parameter for the considered
($p,T$)-range is $S_0 \simeq 0.008 \pm 0.0002$. It indicates on
the nucleated water droplets of the sphere-like form, which is
also confirmed by a visual inspection of snapshots
(Fig.~\ref{fig_snapshot}). We remark here, this result is not the
same with the findings of
Refs.~\cite{Yasuoka/Matsumoto_JCP_1993,Matsubara_JCP_2007}, where
the detected critical clusters in water had the significant
deviations from a spherical form. A possible reason affecting the
observed discrepancy could be different cluster definitions
applied by Matsubara \textit{et al}. in
Ref.~\cite{Matsubara_JCP_2007} and used in the present study
within the statistical treatment.  In addition, the low values for
the nucleated droplets were obtained in
Ref.~\cite{Matsubara_JCP_2007}, $n^*\simeq 16\div22$ partilces,
and the pronounced deviation from a spherical form can be
considered as a signature of the finite size effects: a weak
structural rearrangement in such a system raises the significant
change of its shape.

\section{Conclusions}

The coarse-grained models for particle interactions in molecular
systems provide the good opportunity to study early stages of the
phase transitions by means of the numerical simulations. In this
work, the processes of the steady-state homogeneous
vapor-to-liquid nucleation and the growth of liquid droplets in
water were considered within the mW-model, which treats the
molecular interactions excluding any details of the direct
oxygen-hydrogen interactions and electrostatics. Despite the
apparent coarsening in the description of the molecular
interactions, we have shown that the mW-model provides interesting
information concerning the droplet nucleation in water vapor,
thereby complementing the simulation results obtained earlier
within all-atom models of water such as TIP$4$P and
SPC/E~\cite{Khusnutdinoff/Mokshin_JNCS_2011,Khusnutdinoff/Mokshin_PhysicaA_2012,Yasuoka/Matsumoto_JCP_1993,Matsubara_JCP_2007}.
It is necessary to note, the results reported here are obtained on
the basis of the extended statistical treatment within the MFPT
approach and the thermodynamic integration scheme.

The surface tension of the nucleated droplets was computed within
an approximation that is restricted by the consideration of the
two-particle interactions only without handling the three-particle
contribution to the energy of system. The obtained values
demonstrate the decrease of the surface tension with the
temperature growth. It is necessary to note that the applied
numerical scheme gives the higher values for the liquid-vapor
surface tension of the droplets in comparison with the values for
the liquid-vacuum surface tension of a planar interface reported
in Ref.~\cite{Molinero/Baron,Molinero_JPCB_2009}. We suppose that
this difference is rather a result of the approximations applied
to the surface tension definition than the mW-model product.

Further, the evaluated values of the steady-state nucleation rate
are comparable with the results for all-atom models as well as
with the treatments within the classical nucleation theory.
Unfortunately, we could not to perform the direct comparison of
the obtained nucleation rates with the experimental data, because
we found no experimental $J_s$ for the ($p,T$)-line considered
here. Nevertheless, quantitative extrapolation of the obtained
outcomes indicates on the difference between simulated and
experimental results, that is similar with the known difference
between the experimental data and the CNT
predictions~\cite{Hale_PRL_2010,Viisanen_JCP_1993}. On the other
hand, for the critical size of the nucleated droplet we found the
values within the range $30 \div 100$ particles, which are
expected to be comparable with the experimental data (Fig.~$13$ in
Ref.~\cite{Viisanen_JCP_1993}). So, the additional studies are
highly desirable in this field.

According to our results, the growth of nucleated droplets in the
system is characterized by the remarkable features: the growth law
of the droplet radius follows the power law, $R(t) \propto
t^{1.3}$, and the growth is not steady (with the time-dependent
growth rate $G(t) \propto t^{0.3}$). Moreover, the simple
rescaling on the critical droplet characteristics yields the
unified form of the growth law at all the considered temperatures.

Finally, we found that the critically-sized droplets have a shape,
which is close to spherical one. Note, that the deviations from
spherical shape of the water droplets at homogeneous nucleation,
which were established for the SPC/E-model by Matsubara \textit{et
al.}  (see Ref.~\cite{Matsubara_JCP_2007}), could be simply
originated from the extremely low obtained values for the critical
size $n^*$.

\section{Acknowledgments}

The authors acknowledge B.N.~Hale for helpful correspondence and
R.M.~Khusnutdinoff for many useful discussions.

\bibliographystyle{unsrt}

\begin{thebibliography}{99}

\bibitem{Wolk_JCPB_2001} W\"{o}lk, J.; Strey, R. \textit{J. Phys. Chem. B} \textbf{2001}, \textit{105}, 11683-11701.

\bibitem{Kim_JPCA_2004}
Kim, Y.\;J.; Wyslouzil, B.\;E.; Wilemski, G.; W\"{o}lk, J.; Strey,
R. \textit{J. Phys. Chem. A} \textbf{2004}, \textit{108},
4365-4377.

\bibitem{Manka_JCP_2010} Manka, A.\;A.; Brus, D.; Hyv\"{a}rinen, A.\;-P.;
Lihavainen, H.; W\"{o}lk, J.; Strey, R. \textit{J. Chem. Phys.}
\textbf{2010}, \textit{132}, 244505 1-10.

\bibitem{Brus_JCP_2009} Brus, D.; \v{Z}d\'{i}mal, V.; Uchtmann, H. \textit{J. Chem.
Phys.} \textbf{2009}, \textit{131}, 074507 1-9.

\bibitem{Brus_JCP_2008} Brus, D.; \v{Z}d\'{i}mal, V.; Smol\'{i}k, J. \textit{J. Chem.
Phys.} \textbf{2008}, \textit{129}, 174501 1-8.

\bibitem{Mikheev_JCP_2002}
Mikheev, V.\;B.; Irving, P.\;M.; Laulainen, N.\;S.; Barlow,
S.\;E.; Pervukhin, V.\;V. \textit{J. Chem. Phys.} \textbf{2002},
\textit{116}, 10772-10786.

\bibitem{Luijten_JCP_1997} Luijten, C.\;C.\;M.; Bosschaart, K.\;J.; van Dongen, M.\;E.\;H. \textit{J.
Chem. Phys.} \textbf{1997}, \textit{106}, 8116-8123.

\bibitem{Heist_Exp_water} Heist, R.\;H.; He, H. \textit{J. Phys. Chem. Ref.
Data} \textbf{1994}, \textit{23}, 781-804.

\bibitem{Viisanen_JCP_1993} Viisanen. Y.; Strey, R.; Reiss, H.
\textit{J. Chem. Phys.} \textbf{1993}, \textit{99}, 4680-4692.

\bibitem{Matsubara_JCP_2007} Matsubara, H.; Koishi, T.; Ebisuzaki, T. \textit{J. Chem. Phys.} \textbf{2007}, \textit{127}, 214507 1-11.

\bibitem{Yasuoka/Matsumoto_JCP_1993} Yasuoka, K.; Matsumoto, M. \textit{J. Chem.
Phys.}  \textbf{1998}, \textit{109}, 8451-8462.

\bibitem{Merikanto_JCP_2004} Merikanto, J.; Vehkam\"{a}ki, H.; Zapadinsky, E. \textit{J. Chem.
Phys.} \textbf{2004}, \textit{121}, 914 1-24.

\bibitem{Chen_JCPA_2005} Chen, B.; Siepmann, J.\;I.; Klein, M.\;L. \textit{J. Phys. Chem.
A} \textbf{2005}, \textit{109}, 1137-1145.

\bibitem{Molinero_JPCB_2009} Molinero, V.; Moore, E.\;B. \textit{J. Phys. Chem. B} \textbf{2009}, \textit{113}, 4008-4016.

\bibitem{Moore_Nature_2011}  Moore, E.\;B.; Molinero, V. \textit{Nature} \textbf{2011}, \textit{479}, 506-508.

\bibitem{Reinhardt_JCP_2012} Reinhardt, A.; Doye, J.\;P.\;K. \textit{J. Chem. Phys.} \textbf{2012}, \textit{136}, 054501 1-11.

\bibitem{Bartell_JPCB_2004} Bartell, L.\;S.; Turner, G.\;W. \textit{J. Phys.
Chem. B} \textbf{2004}, \textit{108}, 19742-19747.

\bibitem{Mokshin_JPCM_2007}
Mokshin, A.\;V.; Yulmetyev, R.\;M.; Khusnutdinoff, R.\;M.; Hanggi,
P. \textit{J. Phys.: Cond. Mat.} \textbf{2007}, \textit{19}, 046209
1-16.

\bibitem{Evans_JPA_1972}
Berry, M.\;V.; Durrans, R.\;F.; Evans, R. \textit{J. Phys. A: Gen.
Phys.} \textbf{1972}, \textit{5}, 166-70.

\bibitem{Horse_PRE_2012}
Horsch, M.;  Hasse, H.;  Shchekin, A.\;K.; Agarwal, A.; Eckelsbach,
S.; Vrabec, J.; M\"{u}ller, E.\;A.; Jackson, G. \textit{Phys. Rev.
E.} \textbf{2012}, \textit{85}, 031605 1-12.

\bibitem{Wedekind_JCP_2007} Wedekind, J.; Strey, R.; Reguera, D. \textit{J. Chem. Phys.} \textbf{2007}, \textit{126},
134103 1-7.

\bibitem{Mokshin/Barrat_PRE_2008} Mokshin, A.\;V.; Barrat, J.-L. \textit{Phys. Rev. E} \textbf{2008}, \textit{77}, 021505 1-7.

\bibitem{Mokshin/Barrat_JCP_2009}  Mokshin, A.\;V.; Barrat, J.-L. \textit{J. Chem. Phys.} \textbf{2009}, \textit{130}, 034502 1-6.

\bibitem{J_Frenkel_book_1946} Frenkel, J. \textit{Kinetic Theory of Liquids}; Oxford University Press: London, 1946.

\bibitem{Hansen/McDonald_book_2006} Hansen, J.\;P.; McDonald, I.\;R. \textit{Theory of Simple Liquids}; Academic Press: New York, 2006.

\bibitem{Hanggi_RMP_1990} H{\"a}nggi, P.; Talkner, P.; Borkovec, M. \textit{Rev. Mod. Phys.} \textbf{1990}, \textit{62}, 251-342.

\bibitem{Kashchiev_Nucleation} Kashchiev, D. \textit{Nucleation: Basic
Theory with Applications}; Butterworth Heinemann: Oxford, U.K.,
2000.

\bibitem{Mokshin/Barrat_PRE_2010} Mokshin, A.\;V.; Barrat, J.-L. \textit{Phys. Rev. E} \textbf{2010}, \textit{82}, 021505 1-9.

\bibitem{Stillinger_JCP_1963} Stillinger, F.\;H. \textit{J. Chem. Phys.} \textbf{1963}, \textit{38}, 1486-1494.

\bibitem{Ytreberg_JCP_2006} Ytreberg, F.\;M.; Swendsen, R.\;H.;
Zuckerman, D.\;M. \textit{J. Chem. Phys.} \textbf{2006},
\textit{125}, 184114 1-11.

\bibitem{Khusnutdinoff/Mokshin_PhysicaA_2012} Khusnutdinoff, R.\;M.;
Mokshin, A.\;V. \textit{Physica A} \textbf{2012}, \textit{391},
2842-2847.

\bibitem{Surf_tension_textbook} IAPWS Release on Surface Tension of Ordinary Water Substance, IAPWS, 1994
(http:// www.iapws.org/relguide/surf.pdf).

\bibitem{Molinero/Baron} Baron, R.; Molinero, V. \textit{J. Chem. Theory
Comput} \textbf{2012}, in press, (DOI: 10.1021/ct300121r).

\bibitem{Kiselev/Ely_PhysicaA_2001}
Kiselev, S.\;B.; Ely, J.\;F. \textit{Physica A} \textbf{2001},
\textit{299}, 357-370.

\bibitem{Lu_APL_2006} L\"{u}, Y.\;J.; Wei, B. \textit{Appl. Phys. Lett.} \textbf{2006}, \textit{89},
164106 1-3 and references therein.

\bibitem{Hale_JCP_2005} Hale, B.\;N. \textit{J. Chem. Phys.} \textbf{2005}, \textit{122}, 204509 1-3.

\bibitem{Alexandrov_book_1999}
Alexandrov A.\;A.; Grigoriev B.\;A. \textit{Tables of Thermophysical
Properties of Water and Steam} MEI, Moscow, Russia, 1999.

\bibitem{Tanguy_Review_2011} Rodney, D.; Tanguy, A.; Vandembroucq, D. \textit{Modelling Simul. Mater. Sci. Eng.}
\textbf{2011}, \textit{19}, 083001 1-49.

\bibitem{Khusnutdinoff/Mokshin_JNCS_2011} Khusnutdinoff, R.\;M.;
Mokshin, A.\;V. \textit{J. Non-Cryst. Solids} \textbf{2011},
\textit{357}, 1677-1684.

\bibitem{Hale_PRL_2010} Hale, B.\;N.; Thomason, M. \textit{Phys. Rev. Lett.} \textbf{2010}, \textit{105}, 046101 1-4.

\end{thebibliography}

\end{document}